# Does it Matter Which Citation Tool is Used to Compare the h-index of a Group of Highly Cited Researchers?

[1]Hadi Farhadi, [2]Hadi Salehi, [3]Melor Md Yunus, [4]Arezoo Aghaei Chadegani, [4]Maryam Farhadi, [4]Masood Fooladi and [5]Nader Ale Ebrahim

[1]School of Psychology and Human Development, Faculty of Social Sciences and Humanities, Universiti Kebangsaan Malaysia (UKM), Malaysia
[2]Faculty of Literature and Humanities, Najafabad Branch, Islamic Azad University, Najafabad, Isfahan, Iran
[3]Faculty of Education, Universiti Kebangsaan Malaysia (UKM), Malaysia
[4]Department of Accounting, Mobarakeh Branch, Islamic Azad University, Mobarakeh, Isfahan, Iran
[5]Department of Engineering Design and Manufacture, Faculty of Engineering, University of Malaya, Kuala Lumpur, Malaysia

**Abstract:** *h-index* retrieved by citation indexes (Scopus, Google scholar, and Web of Science) is used to measure the scientific performance and the research impact studies based on the number of publications and citations of a scientist. It also is easily available and may be used for performance measures of scientists, and for recruitment decisions. The aim of this study is to investigate the difference between the outputs and results from these three citation databases namely Scopus, Google Scholar, and Web of Science based upon the h-index of a group of highly cited researchers (Nobel Prize winner scientist). The purposive sampling method was adopted to collect the required data. The results showed that there is a significant difference in the h-index between three citation indexes of Scopus, Google scholar, and Web of Science; the Google scholar h-index was more than the h-index in two other databases. It was also concluded that there is a significant positive relationship between h-indices based on Google scholar and Scopus. The citation indexes of Scopus, Google scholar, and Web of Science may be useful for evaluating h-index of scientists but they have some limitations as well.

**Key words:** h-index, Scopus, Google Scholar, Web of Science, Nobel Prize, Physics, Chemistry, Economic Sciences.

## INTRODUCTION

Nowadays, when citation information is needed there are a few sources to rely on. The most comprehensive web version of citation indexes sources are Scopus (http://www.scopus.com/), Google Scholar (http://scholar.google.com/), and the Web of Science (http://portal.isiknowledge.com/). Scopus, Google scholar, and Web of Science are regarded as the most useful and trustful source for searching. They are valuable tools for searching which provide the citation searching and ranking by times cited (Mikki, 2009). These three sources of citation indexes are multidisciplinary and international coverage and used to assess the scientific output worldwide (Bar-Ilan, 2008).

*Background:*
To quantify the research output of a single scientist, Hirsch (2005) introduced the h-index. *h-index* is calculated by the number of publications and the number of citations received. As defined by Hirsch (2005), a scientist has index h if h of his or her $N_p$ papers have at least h citations each and the other ($N_p$-h) papers have $\leq$ h citations each. It is a single number which supports a good representation of the scientific lifetime achievement (Egghe & Rousseau, 2006; van Raan, 2006). For example, if a scientist has published 15 papers that each had at least 15 citations, h-index of him/her will be 15. Although this single number is simple to compute and it takes into account both the quantity and impact of the papers, it is essential to consider which database is reporting this single number (h-index).

*h-index* is computed based upon the data from the aforementioned citation indexes to measure the scientific output of a researcher. The data from these citation indexes are important for scientists and universities in all over the world. For instance, h-index is one of the impressive and effective factors used by promotion committees to evaluate research productivity and impact at universities. Although, citing and cited documents from these citation tools are associated to each other, h-index can be obtained separately from different citation databases of Scopus, Google Scholar, and Web of Science. That is because each of these citation indexes has a

**Corresponding Author:** Hadi Farhadi (School of Psychology and Human Development, Faculty of Social Sciences and Humanities, Universiti Kebangsaan Malaysia (UKM), Malaysia
E-mail: farhadihadi@yahoo.com

different collection policy which influences both the publications covered and the number of citations (Bar-Ilan, 2008). Hence, the outputs and the results from these sources of citation indexes should not be approximately similar.

This study aims to distinguish the major differences on h-index of scientists between three citation databases of Scopus, Google Scholar, and Web of Science. Therefore, based upon the literature, this study attempts to answer the following research questions:
1) How much the results from these sources of citation indexes are different?
2) Can universes and promotion committees use the h-index of a scientist as a tool for synthesizing and depicting the scientific production in a more exhaustive way?

To answer the above research questions, the current study investigated the differences between the outputs and the results from these three citation databases (Scopus, Google Scholar, and Web of Science) based upon the h-index of a group of highly cited researchers (Nobel Prize winner scientists).

*Methodology:*

There exist three commonly web versions of citation indexes sources to measure the h-index of a scientist. Hence, this study computed the h-index for 12 Nobel Prize winners based upon data from Scopus, Google Scholar, and Web of Science for three different fields of study namely physics, chemistry and economic sciences. Then, we compared the h-index data of these 12 Nobel Prize winners to find out the correlation between the h-index of three mentioned citation indexes sources. Purposive sampling method was used to collect the required data. To do so, we used the official web site of the Nobel Prize (http://www.nobelprize.org/) as our starting point, which lists the all Nobel Prize winners.

Every year, the Nobel Foundation in Stockholm, Sweden, presents an international award called "Nobel Prize" to those who have made outstanding contributions to their disciplines. The Nobel Prize includes a medal, a diploma, and cash award given annually to those who during the preceding year have given the greatest benefit to human beings. Every year, the respective Nobel Committees invite thousands of university professors, members of academies and scientists from different countries, previous Nobel laureates, and members of parliamentary assemblies to submit candidates for the Nobel Prizes for the coming year. These nominators are chosen in such a way that as many countries and universities as possible are represented over time.

For this study, we selected 12 scientists from the official website of the Nobel Prize who have won the Nobel Prize in three different fields of physics, chemistry, and economic sciences. Then the name of each scientist was searched in three citation indexes sources of Scopus, Google Scholar, and Web of Science (see Figures 1, 2 and 3). We brought Serge Haroche (the physics Nobel prize 2012 winner) as an example in following figures. As can be seen in the following figures, the h-index of Serge Haroche is 34, 35 and 21 based upon Scopus, Google Scholar, and Web of Science, respectively.

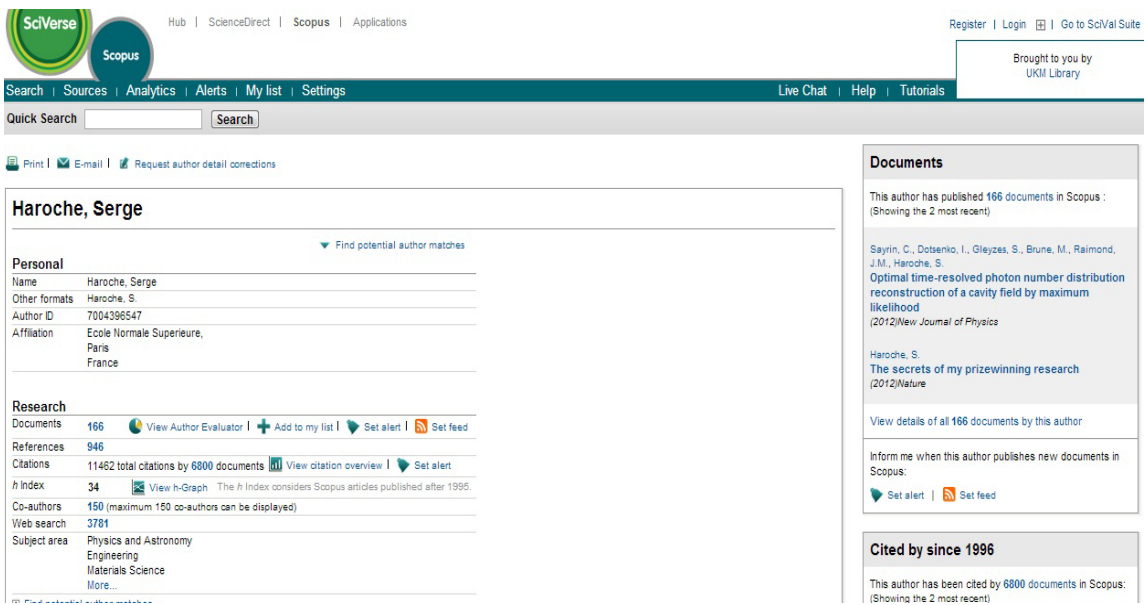

**Fig. 1:** Serge Haroche's h-index by Scopus is 34.

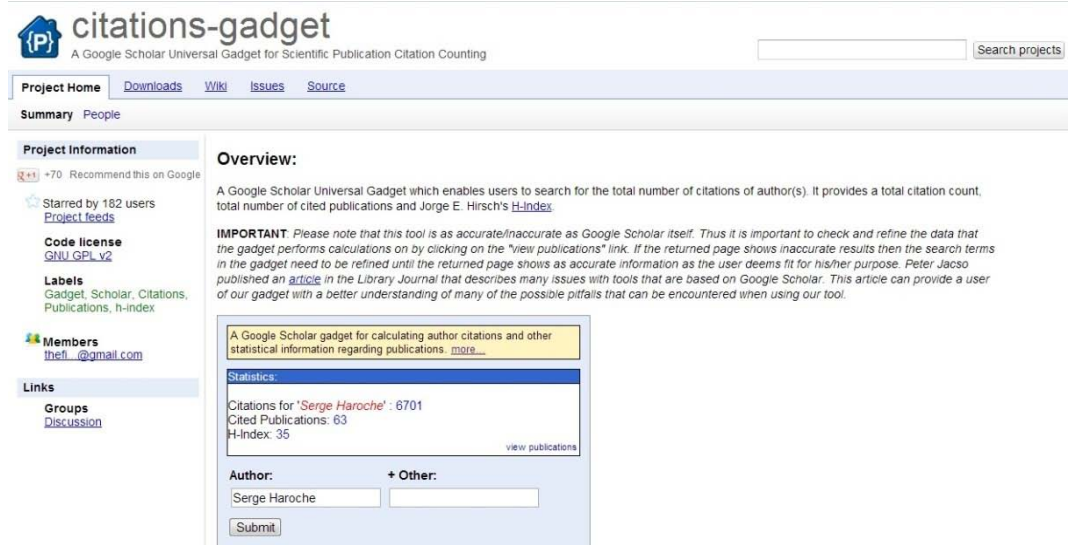

**Fig. 2:** Serge Haroche's h-index by Google Scholar is 35.

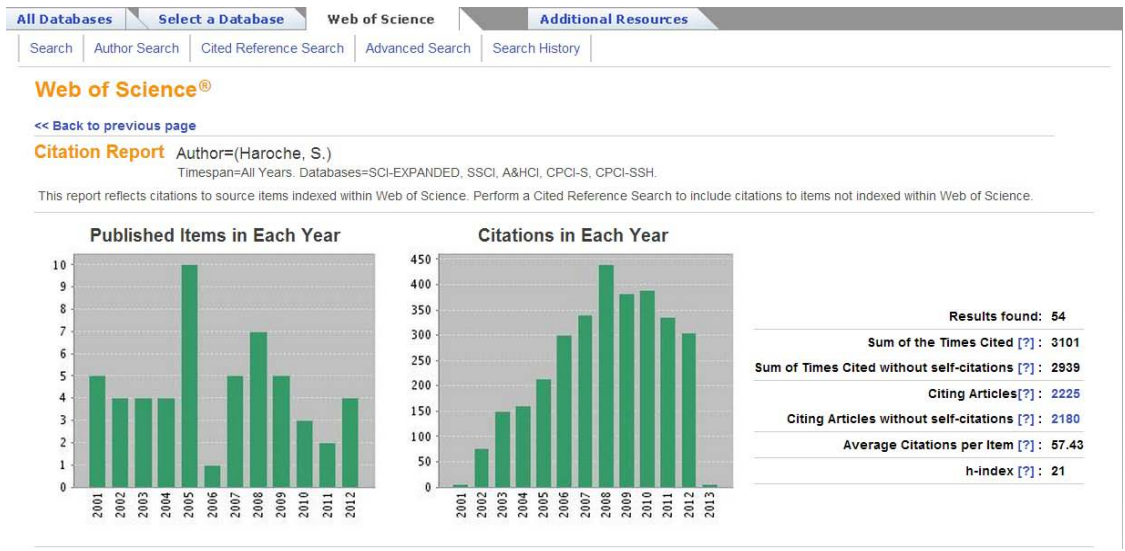

**Fig. 3:** Serge Haroche's h-index by Web of Science is 21.

## RESULTS AND DISCUSSION

Table 1 shows the h-index of 12 Nobel Prize winner scientists in three different citation indexes of Scopus, Google Scholar and Web of Science.

**Table 1:** h-index of 12 Nobel Prize winner scientists in three different citation indexes.

| Noble Prize Winner's Name | Year | Research field | Web of Science | Scopus | Google Scholar |
|---|---|---|---|---|---|
| Serge Haroche | 2012 | Physics | 21 | 34 | 35 |
| David J. Wineland | 2012 | Physics | 20 | 47 | 23 |
| Saul Perlmutter | 2011 | Physics | 35 | 38 | 32 |
| Brian P. Schmidt | 2011 | Physics | 21 | 46 | 62 |
| Robert J. Lefkowitz | 2012 | Chemistry | 31 | 106 | 167 |
| Brian K. Kobilka | 2012 | Chemistry | 24 | 63 | 70 |
| Dan Shechtman | 2011 | Chemistry | 11 | 5 | 13 |
| Akira Suzuki | 2010 | Chemistry | 85 | 56 | 79 |
| Alvin E. Roth | 2012 | Economic Sciences | 4 | 28 | 68 |
| Thomas J. Sargent | 2011 | Economic Sciences | 11 | 21 | 77 |
| Christopher A. Sims | 2011 | Economic Sciences | 13 | 13 | 64 |
| Peter A. Diamond | 2010 | Economic Sciences | 6 | 14 | 61 |

Our first step in understanding the differences between Scopus, Google Scholar, and Web of Science based upon the scientists' h-index was to examine the means of each citation tool resource. Figure 4 shows the mean of h-index for Nobel Prize winner scientists based upon data from Google Scholar (mean=62.58), Scopus (39.25), and Web of Science (23.5). As can be seen in Figure 4, the mean of Google scholar h-index (mean = 62.58) for scientist in this research is more than the mean of Scopus h-index (mean = 39.25) and mean of Web of Science h-index (mean = 23.5). The results can be reasonable because Google Scholar covers more articles indexed than other two citation tools and it contains books and conference proceedings which may alter considerable citation metrics (Vanclay, 2007; Bar-Ilan, 2008; Mikki, 2009).

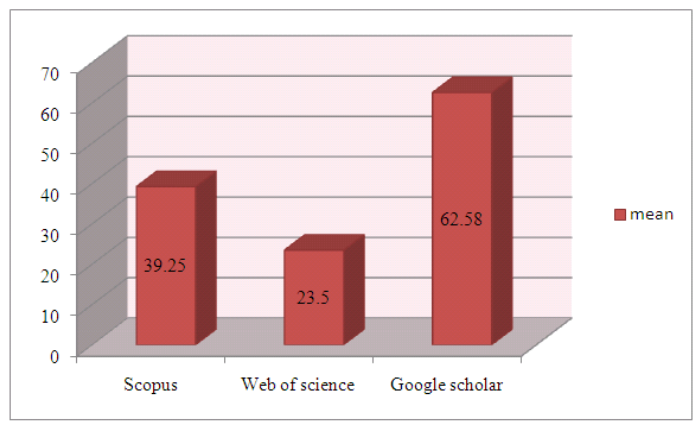

**Fig. 4:** Comparing the h-index of 12 Nobel Prize winners based upon data from Scopus, Google Scholar and Web of Science.

Table 2 also shows the multiple regression between the h-index of 12 Nobel Prize winners based upon data from Scopus, Google Scholar and Web of Science. The multiple regression results show that there is a positive significant relationship between Scopus and Google Scholar h-index ($\beta=.799$, $p<.05$). This indicates that more h-index in Scopus citation tool corresponds with more Google Scholar h-index. In other words, we can conclude that the h-index of Scopus can predict the Google scholar h-index. However, this study could not find a significant relationship between the h-index in Web of Science and Google Scholar h-index.

**Table 2:** Multiple regression between the h-index of 12 Nobel Prize winners based upon data from Scopus, Google Scholar and Web of Science.

|  | B | Beta | t |
|---|---|---|---|
| Constant | 24.761 |  | 1.561 |
| Web of Science | -.300 | -.164 | -.639 |
| Scopus | 1.143 | .799 | 3.122* |
| $R^2= 0.54$, $F= 5.32$, $p<0.05$ |  |  |  |

Based upon the samples' field of studies, the results showed that the h-index mean of chemistry (mean = 59.17) is more than the h-index mean of physics (mean = 37.5) and economy (mean = 31.67). It means that the citation and publication in chemistry area are more than those of physics and economy (see Figure 5).

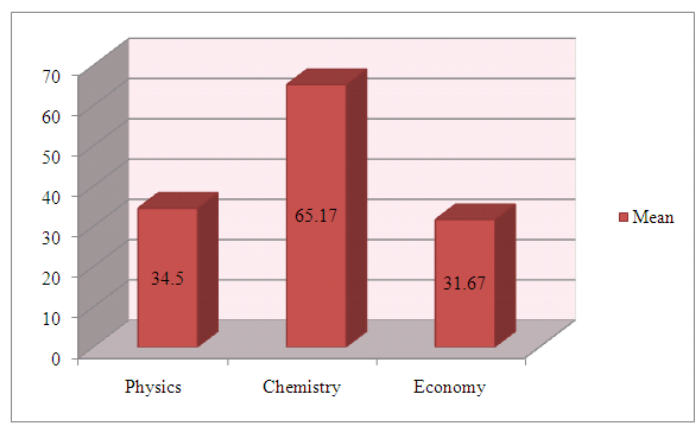

**Fig. 5:** Comparing the h-index of 12 Nobel Prize winners based upon their research area.

*Conclusions:*

h-index retrieved by citation indexes of Scopus, Google Scholar, and Web of Science is used to measure the scientific performance and the research impact studies based upon the number of publications and citations of a scientist. It also is easily available and may be used for performance measures of scientists and for recruitment decisions. As the results showed for the samples of this study, the difference in the h-index between three citation indexes of Scopus, Google Scholar and Web of Science is obvious and noticeable. The Google Scholar h-index is more in comparison with the h-index in two other databases. Moreover, the findings showed that there is a significant relationship between h-indices based upon Google Scholar and Scopus.

We can conclude that it matters which citation tools are used to compute the h-index of scientists. Therefore, the universities and promotion committees need to consider which citation tool is used to compute the h-index of a researcher or a lecturer. In general, when we intend to report the h-index of a scientist, we need to consider and mention the sources of citation indexes. We cannot compare the h-index of two scientists without mentioning the sources of citation indexes. The citation indexes of Scopus, Google Scholar and Web of Science might be useful for evaluating the h-index of scientists but they have their own limitations as well.